\documentclass[9pt,twocolumn,twoside]{opticajnl}
\journal{opticajournal} 

\setboolean{shortarticle}{true}

\usepackage{amsmath,amssymb,amsfonts}
\usepackage{graphicx}
\usepackage{textcomp}
\usepackage{xcolor}
\usepackage{multirow}
\usepackage{booktabs}
\usepackage{array}
\usepackage{lineno}

\title{Per-Channel Launch-Power Optimization in Hollow-Core Fiber Systems}
\author[1,*]{Md Ghulam Saber}
\author[1]{Qingyi Guo}
\author[1]{Zhiping Jiang}

\affil[1]{Ottawa Research Center, Huawei Technologies Canada, 303 Terry Fox Drive, Kanata, ON, K2K 3J1, Canada.}

\affil[*]{md.ghulam.saber@huawei.com}

\begin{abstract}
In single-mode fiber (SMF) the Kerr effect ties every channel's quality of
transmission to its neighbors' launch powers through cross-phase modulation (XPM), four-wave
mixing (FWM) and inter-channel stimulated Raman scattering (ISRS), forcing a jointly planned launch
profile. Hollow-core fiber (HCF), with a Kerr coefficient three to four orders of
magnitude below silica, turns the per-channel powers into nearly independent knobs
limited only by the shared amplifier budget. We build a per-channel generalized
signal-to-noise ratio (GSNR) budget for amplified HCF links, including amplified
spontaneous emission (ASE) with a wavelength- and output-power-dependent erbium-doped
fiber amplifier (EDFA) noise figure (NF), inter-modal interference (IMI), nonlinearity in amplifier
pigtails, CO$_2$ gas-line loss and a flat transceiver (TRx) noise ceiling, and
derive a sensitivity law that predicts when power shaping pays: its gain is bounded by
the ASE noise share, canceled by self-phase modulation (SPM) at the SMF single-channel
optimum, and positive in HCF. Across 80$\times$64-GBaud C-band links over 400--3200 km,
per-channel optimization buys up to 1.0 dB of worst-channel GSNR over a flat launch
as the EDFA NF spread grows to 4 dB, cuts cross-channel power
sensitivity by more than two orders of magnitude relative to SMF, and reaches a
given GSNR at about $3$ dB lower aggregate amplifier output. At a fixed
amplifier budget this becomes a $1.26$ to $1.41$ times worst-channel reach extension,
against at most $15\%$ on the nonlinearity-capped SMF link. 
\end{abstract}

\setboolean{displaycopyright}{false} 

\begin{document}

	\maketitle
	\textbf{Introduction.}
	Launch-power optimization in wavelength-division multiplexed (WDM) systems is strongly constrained by fiber nonlinearity. In SMF, the Kerr effect generates nonlinear interference (NLI): SPM limits each channel individually, while its cross-channel terms XPM and FWM, with ISRS, tie each channel to its neighbors' launch powers \cite{Poggiolini2012}. The profile must therefore be planned jointly, e.g., via the local-optimization/global-optimization (LOGO) strategy with pre-emphasis and tilt \cite{Poggiolini2013_LOGON,Ives2014_power}.

	HCF largely removes this constraint: nested/double-nested antiresonant nodeless fibers (NANFs/DNANFs) now reach attenuation at or below that of SMF \cite{Petrovich2025,saber2026_perspective} with a Kerr coefficient three to four orders of magnitude lower \cite{Fokoua2023}, making XPM, FWM, ISRS and Brillouin scattering negligible and supporting launch powers well above SMF practice \cite{Hong2024_HCF_terabit,Sohanpal2026_launch}. The per-channel powers thus become nearly independent knobs under a shared amplifier budget. This matters because practical systems carry wavelength-dependent GSNR deficits: amplifier noise-figure variation \cite{Becker1999_EDFA,Amonics_AEDFA} and frequency-selective CO$_2$ absorption \cite{Wang2025_OL_CO2,Sillekens2026_GLA}. Hong \textit{et~al.} \cite{Hong2026_TPO} showed this experimentally, adding up to ${\sim}6$~dB on gas-impaired channels to equalize the pre-forward-error-correction (FEC) bit-error ratio (BER) across 32 real-time 800G channels over 442.7~km. This work differs in constraint, objective, impairments and scope. \cite{Hong2026_TPO} holds the amplifier output fixed and redistributes it; we make the total output the optimized resource, which is what turns the result into statements about amplifier sizing and reach. The objective is max-min GSNR rather than uniform BER; the model carries a wavelength- and output-dependent NF, a TRx ceiling, IMI/multipath interference (MPI) and pigtail NLI alongside gas, with an SMF baseline solved on the same footing; and the study spans 5--40 spans and 80 channels with an analytical sensitivity law. We read \cite{Hong2026_TPO} as evidence that reallocation of this magnitude is realizable on deployed hardware, and the results below as the consequences a fixed-output gas-only experiment cannot expose.

	The model below adds ASE with a wavelength- and output-power-dependent EDFA NF, distributed and discrete IMI, residual Kerr nonlinearity and a flat TRx noise ceiling. Sweeps of NF spread, gas loss, budget and reach give up to 1.0~dB worst-channel improvement, approximately $3$~dB lower aggregate output and a $1.26$ to $1.41\times$ fixed-budget reach extension.

	\textbf{System Model.}
	\label{sec:model}
	We consider $N=80$ dual-polarization channels on a 75-GHz grid at a symbol rate
	$R_s=64$~GBaud (6~THz occupied) over $N_s$ spans of $L_s=80$~km with lumped
	EDFAs that restore each channel to its launch power $p_i$ at every span input.
	The baseline uses $N_s=10$ (800~km); the reach study sweeps $N_s=5$--$40$
	(400--3200~km) for the GSNR-versus-distance curves and extends to $N_s=200$
	(16\,000~km) where a reach threshold must be located.
	Table~\ref{tab:params} lists the fiber parameters, taken from
	\cite{saber2026_qamlimits}, followed by the system parameters. The C-band grid starts at 1520~nm; a gas-line
	scenario uses an L-band grid starting at 1572~nm. We assume soft-decision FEC at
	a pre-FEC bit-error ratio of $2\times10^{-2}$ with the probabilistically shaped
	(PS) quadrature amplitude modulation (QAM) thresholds of \cite{saber2026_qamlimits}, referencing the PS-64-QAM
	requirement (18.6~dB with 1-dB margin) and noting PS-256-QAM (23.9~dB) where
	approached.

	\begin{table}[!t]
		\caption{System and Fiber Parameters. DMA: differential modal attenuation.}
		\label{tab:params}
		\centering
		\footnotesize
		\setlength{\tabcolsep}{3.5pt}
		\resizebox{\columnwidth}{!}{%
			\begin{tabular}{lcc}
				\toprule
				\textbf{Parameter} & \textbf{SMF} & \textbf{HCF (NANF/DNANF)} \\
				\midrule
				Attenuation (dB/km) & 0.20 & 0.11 \\
				Dispersion (ps/nm/km) & 17.0 & 3.0 \\
				$\gamma$ (W$^{-1}$km$^{-1}$) & 1.4 & 0.001 \\
				Distributed IMI $\kappa$ (dB/km) & --- & $-73$ (baseline); $-55$ \\
				Extra span loss (dB) & 0.5 & 2.0 (splices, mode-field adapters) \\
				ISRS slope $C_r$ (W$^{-1}$km$^{-1}$THz$^{-1}$) & 0.028 & $\approx 0$ \\
				HCF segment / DMA & --- & 5~km / 10~dB/km \\
				LP$_{11}$ excitation: splice, junction (dB) & --- & $-30$, $-35$ \\
				\midrule
				Channels / grid / rate & \multicolumn{2}{c}{80 / 75~GHz / 64~GBaud} \\
				EDFA NF (min, spread) & \multicolumn{2}{c}{4.5~dB, 0--4~dB across band} \\
				NF power penalty & \multicolumn{2}{c}{$+0.42$~dB/dB above 37~dBm \cite{Amonics_AEDFA}} \\
				TRx back-to-back SNR at 64~GBaud & \multicolumn{2}{c}{25.3~dB, flat across the band} \\
				Amplifier output budget & \multicolumn{2}{c}{21--43~dBm total (26~dBm cap for SMF)} \\
				SMF pigtail per EDFA & \multicolumn{2}{c}{5~m (both sides)} \\
				\bottomrule
		\end{tabular}}
	\vspace{-.5cm}
	\end{table}

	\emph{Per-Channel GSNR Budget.}
	Collecting additive noise powers referred to the receiver in the channel
	bandwidth, the per-channel GSNR obeys
	\begin{equation}
		\frac{1}{\mathrm{SNR}_i(\mathbf{p})} =
		\underbrace{\frac{a_i}{p_i}}_{\text{ASE}}
		+ \underbrace{b_i}_{\text{IMI/MPI}}
		+ \underbrace{c_{ii}\,p_i^2 + \!\sum_{j\neq i} c_{ij}\,p_j^2}_{\text{Kerr NLI}}
		+ \underbrace{\frac{1}{T_i}}_{\text{TRx}},
		\label{eq:budget}
	\end{equation}
	where $\mathbf{p}=(p_1,\dots,p_N)$ is the launch-power vector, $a_i$ (W) the
	accumulated ASE power [\eqref{eq:ase}], $b_i$ the power-neutral
	interference-to-signal fraction from IMI and MPI,
	$c_{ij}$ (W$^{-2}$) the Kerr NLI kernels and $T_i$ the linear-scale TRx
	back-to-back (B2B) SNR ceiling. This polynomial form is standard in
	Gaussian-noise (GN) launch-power optimization \cite{Ives2014_power}; $1/T_i$
	follows \cite{Galdino2017_trxnoise,Klaus2022_HCF} and $b_i$ extends it to HCF
	\cite{saber2026_qamlimits}. The ASE coefficient follows the cascaded-EDFA
	accumulation \cite{Becker1999_EDFA,Klaus2022_HCF},
	\begin{equation}
		a_i = N_s\, h\nu_i R_s\, F_i(\lambda, P_{\mathrm{out}})\, \big(G_i - 1\big),
		\quad G_i = 10^{\frac{(\alpha + g_i)L_s + L_x}{10}},
		\label{eq:ase}
	\end{equation}
	with $h$ Planck's constant, $\nu_i$ the channel center frequency, $R_s$ the
	noise reference bandwidth, $F_i$ the linear-scale EDFA NF, $G_i$ the span gain
	restoring the launch power, $\alpha$ the attenuation (dB/km), $g_i$ the CO$_2$
	excess loss (dB/km) and $L_x$ the lumped splice/connector loss per span (dB).
	Equation~\ref{eq:ase} carries the three mechanisms
	power shaping can address: the wavelength-dependent NF $F_i$, a band-independent
	penalty added to $F_i$ when the total output $P_{\mathrm{out}}=\sum_i p_i$
	exceeds the high-power knee, i.e.\ the output above which the amplifier can no
	longer hold its rated inversion and its NF rises with output power. The knee is
	at 37~dBm with a $+0.42$~dB/dB slope, linearized across the 37- and 43-dBm
	members of one commercial EDFA family \cite{Amonics_AEDFA}.
	The gas loss is $g_i$. Following
	\cite{saber2026_qamlimits}, the CO$_2$ spectrum is a comb of pressure-broadened
	Lorentzian lines (1.5-GHz full width at half-maximum, ${\sim}23$-GHz spacing)
	scaled so that the loss at a line center equals the up-to-0.5-dB/km peak value
	reported in \cite{Wang2025_OL_CO2}; averaged over $R_s$, $g_i$ is at most
	${\approx}0.05$~dB/km for a 64-GBaud channel on a line. The power-neutral
	fraction $b_i$ holds the distributed IMI term $\kappa L_{\mathrm{tot}}$
	($L_{\mathrm{tot}}=N_sL_s$) \cite{Poggiolini2022_HCF} and the all-pairs splice
	multipath sum with DMA
	\cite{Mlejnek2015,saber2026_qamlimits}. The Kerr kernels $c_{ij}$ use GN-model
	closed forms \cite{Poggiolini2012,Semrau2019_ISRS} for the fiber and the 5-m SMF
	pigtails at each amplifier; for SMF an ISRS per-span tilt (re-flattened by the
	EDFA) adds a second coupling path absent in HCF.
	$T_i$ is the TRx B2B SNR ceiling, anchored to the 64-GBaud value of
	\cite{saber2026_qamlimits} and held flat across the band
	\cite{Emmerich2023_rxchar}. The gas-notch
	intersymbol interference (ISI) penalty is a waveform distortion, not an added
	noise power: following \cite{saber2026_qamlimits} it adds
	$\delta_{\mathrm{ISI}}=\delta_{\max}(1-e^{-g_iL_{\mathrm{tot}}/A_0})$~dB to the
	required SNR, taking the accumulated in-band loss $g_iL_{\mathrm{tot}}$ as the
	notch-depth proxy, built from the same channel-bandwidth-averaged $g_i$ that
	enters \eqref{eq:ase}. Residual ISI grows
	with notch depth while the notch is shallow and saturates at $\delta_{\max}$ once
	a bounded-tap equalizer can no longer invert it. The constants
	$\delta_{\max}{\approx}5.5$~dB and $A_0{\approx}6$~dB are anchored to the
	equalization and pre-equalization results of \cite{Sillekens2026_GLA} and
	calibrated so that the worst L-band channel on a line approaches the
	${\sim}6$-dB penalty reported in \cite{Wang2025_OL_CO2}. Being a deterministic
	waveform distortion, $\delta_{\mathrm{ISI}}$ raises the SNR a channel needs
	to close its FEC target: it changes which modulation format is feasible, but
	leaves the GSNR of \eqref{eq:budget} unchanged.

		\begin{figure}[!t]
		\centering
		\includegraphics[width=.999\columnwidth]{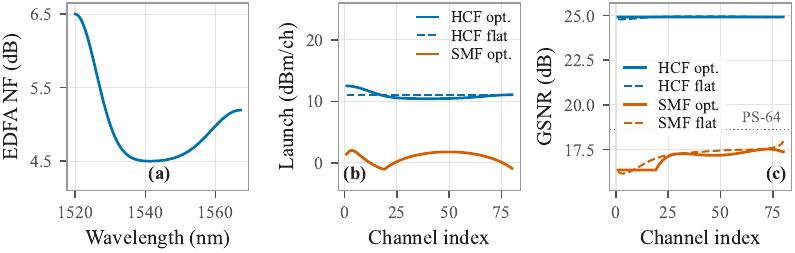}
		\vspace{-.2cm}
		\caption{Baseline inputs and C-band performance (800~km, 30 dBm):
			(a)~EDFA NF; (b)~launch profiles; and (c)~per-channel GSNR
			(dashed: flat loading).}
		\label{fig:baseline}
		\vspace{-.38cm}
	\end{figure}
	
		\begin{figure}[!t]
		\centering
		\includegraphics[width=.96\columnwidth]{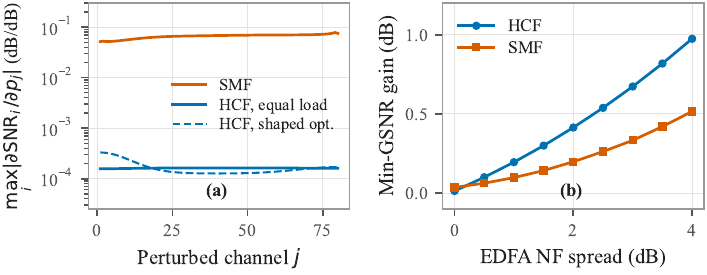}
		\vspace{-.2cm}
		\caption{Decoupling and deficit scaling: (a)~largest cross-channel GSNR
			change per dB on channel $j$ (800~km, 30 dBm); (b)~worst-channel gain vs
			EDFA NF spread.}
		\label{fig:decoupling}
		\vspace{-.45cm}
	\end{figure}

	\emph{Power-Scaling Taxonomy and Sensitivity Law.}
	\label{subsec:sensitivity}
	Reading the terms of \eqref{eq:budget} as separate ratios,
	$\mathrm{SNR}_{\mathrm{ASE},i}=p_i/a_i$, $\mathrm{SNR}_{\mathrm{IMI},i}=1/b_i$,
	$\mathrm{SNR}_{\mathrm{TRx},i}=T_i$ and $\mathrm{SNR}_{\mathrm{NLI},i}=
	[c_{ii}p_i^2+\sum_{j\neq i}c_{ij}p_j^2]^{-1}$, only
	$\mathrm{SNR}_{\mathrm{ASE},i}$ improves with power, in proportion to $p_i$.
	$\mathrm{SNR}_{\mathrm{IMI},i}$ and $\mathrm{SNR}_{\mathrm{TRx},i}$ are
	independent of $p_i$ because their noise powers scale with the signal; we call
	them power-neutral. $\mathrm{SNR}_{\mathrm{NLI},i}$ degrades as $p_i^{-2}$ and
	its cross terms $c_{ij}$ tie channel $i$ to its neighbors' powers, which is
	what forces the SMF profile to be planned jointly. Writing the total noise of
	channel $i$ as $N_i(p_i) = a_i + \tilde{b}_i\,p_i + c_{ii}\,p_i^3$
	($\tilde{b}_i$ collecting the linear-in-power terms), $\mathrm{SNR}_i = p_i/N_i$
	gives the self-sensitivity
	\begin{equation}
		S_i \equiv \frac{\partial\,\mathrm{SNR}_i[\mathrm{dB}]}{\partial\,p_i[\mathrm{dB}]}
		= \rho_{\mathrm{ASE},i} - 2\rho_{\mathrm{SPM},i},
		\label{eq:sensitivity}
	\end{equation}
	where $\rho_{x,i}$ is the share of term $x$ in channel $i$'s total noise:
	the linear-in-power terms drop out, ASE contributes $+1$, the cubic SPM term
	$-2$. Only SPM appears: $\sum_{j\neq i}c_{ij}p_j^2$ does not depend on $p_i$,
	so XPM cancels from the self-derivative and acts only through $K_{ij}$. At the
	GN-model single-channel optimum $p_i^\star=(a_i/2c_{ii})^{1/3}$
	\cite{Poggiolini2012,Ives2014_power} ASE power is exactly twice SPM power, so
	$\rho_{\mathrm{ASE}}=2\rho_{\mathrm{SPM}}$ and $S_i=0$: in SMF the marginal
	self-benefit of power vanishes. \eqref{eq:sensitivity} assumes fixed
	$a_i$; SMF ISRS makes $a_i$ profile-dependent, a term we evaluate numerically.
	The cross-sensitivity
	$K_{ij}\equiv\partial\,\mathrm{SNR}_i[\mathrm{dB}]/\partial\,p_j[\mathrm{dB}]
	=-2\rho_{\mathrm{XPM},ij}<0$ makes boosting a channel a negative-sum game, the
	analytical reason per-channel refinement beyond LOGO yields ${<}0.1$~dB in an
	80-channel C-band link \cite{Ives2014_power}. In HCF $\rho_{\mathrm{SPM}}\approx0$,
	so $S_i\approx\rho_{\mathrm{ASE},i}>0$ and $K_{ij}\approx0$: raising one channel's
	power improves it in proportion to its ASE share and leaves every other channel
	unchanged. The same expression caps the gain at
	${\approx}\rho_{\mathrm{ASE},i}\,\Delta_i$, where the spectral deficit $\Delta_i$
	is the GSNR shortfall of channel $i$ relative to the best channel under flat
	loading, arising here from the wavelength dependence of the EDFA NF and from gas
	absorption. The gain is largest where ASE dominates (long reach, tight budgets,
	gas channels) and compressed where the flat TRx ceiling or the IMI floor does. We solve the max-min GSNR problem on
	$\sum_i p_i \le P_{\mathrm{tot}}$: for HCF it decouples ($\mathrm{SNR}_i$ is
	monotone below the very high $p_i^\star$) and is solved by bisection on the GSNR
	target; the coupled SMF link uses Sequential
	Least Squares Programming (SLSQP).
	Each solve holds $P_{\mathrm{tot}}$ fixed and evaluates $F_i$ there, so $F_i$ is
	constant within it and $\partial F_i/\partial P_{\mathrm{out}}$ never enters
	\eqref{eq:sensitivity}; the amplifier-budget sweep below is an outer scan
	over $P_{\mathrm{tot}}$. The budget binds below the knee; above it the
	optimum stops at the per-channel GN point and leaves the rest unspent, up to
	4.5~dB at a 43-dBm budget, so $F_i$ evaluated at $P_{\mathrm{tot}}$ is an upper
	bound there. Letting $F_i$ track
	the achieved $\sum_i p_i$ adds a coupling of ${\approx}1\times10^{-4}$~dB/dB, zero below
	the knee and three orders of magnitude under SMF. Gains are referenced to a flat
	profile at equal total power, $p_i$ otherwise unconstrained.

	\textbf{Results.}
	\label{sec:results}

	\emph{Baseline: Where the Shaping Gain Goes.}
	\label{subsec:res_baseline}
	Figure~\ref{fig:baseline}(a) shows the only modeled spectral deficit: a 2-dB NF
	spread peaking at the blue edge, the canonical three-level-inversion behavior
	\cite{Becker1999_EDFA,Amonics_AEDFA}. Figure~\ref{fig:baseline}(b),(c) gives the launch
	and GSNR spectra at 800~km, 30~dBm. The HCF optimizer tilts about 2~dB toward
	the blue-edge channels where the NF peaks, lifting the worst channel from 24.79
	to 24.91~dB while the band-averaged GSNR is unchanged to 0.01~dB. The coupled
	SMF max-min optimum gains 0.20~dB on its worst channel, but only by jointly
	re-planning all 80 powers:
	changing any one SMF channel perturbs its neighbors' GSNR ${\sim}27$~dB more
	strongly than in HCF (Fig.~\ref{fig:decoupling}(a)). Boosting one SMF channel alone is
	self-limiting, as \eqref{eq:sensitivity} requires: $S_i$ falls to zero within a
	couple of dB of boost and turns negative beyond. SMF also sits ${\sim}8$~dB below
	HCF in absolute GSNR, equalizing its blue edge at a 16.4-dB floor
	(Fig.~\ref{fig:baseline}(c)).

	A single-channel power-perturbation calculation illustrates the decoupling:
	boosting the worst HCF channel by $+3$~dB improves it $+0.15$~dB and costs every
	other channel $<0.001$~dB; in SMF the same boost gains $+0.16$~dB but costs a
	neighbor $-0.30$~dB.
	Figure~\ref{fig:decoupling}(a) generalizes this: $\max_{i\neq j}|\partial
	\mathrm{SNR}_i/\partial p_j|$ is ${\sim}0.07$~dB/dB in SMF, where a 3-dB
	adjustment moves a neighbor by ${\sim}0.2$~dB, comparable to operating margins,
	versus $1.6\times10^{-4}$~dB/dB in HCF, 27~dB lower. The residual HCF coupling is
	XPM in the amplifier pigtails, not the HCF, so it grows with power: at the
	shaped optimum it rises to $3.3\times10^{-4}$~dB/dB (dashed), still ${\sim}24$~dB
	below SMF.
		\begin{figure}[!t]
		\centering
		\includegraphics[width=.6\columnwidth]{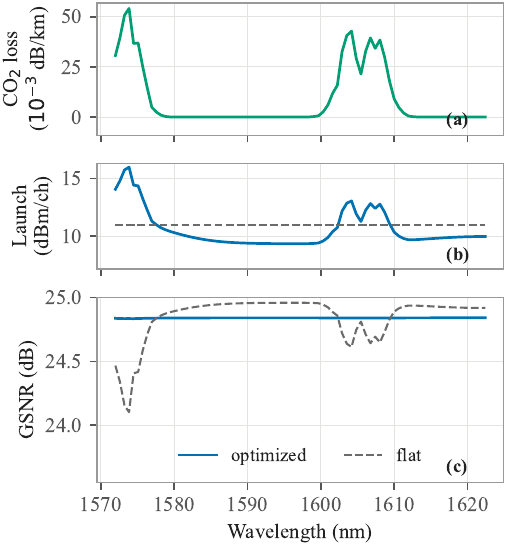}
		\vspace{-.2cm}
		\caption{L-band CO$_2$ scenario: (a) channel-averaged excess loss at
			64~GBaud; (b) optimized versus flat launch; and (c) GSNR.}
		\label{fig:gas}
		\vspace{-.6cm}
	\end{figure}
	\begin{figure*}[!t]
		\centering
		\includegraphics[width=.79\textwidth]{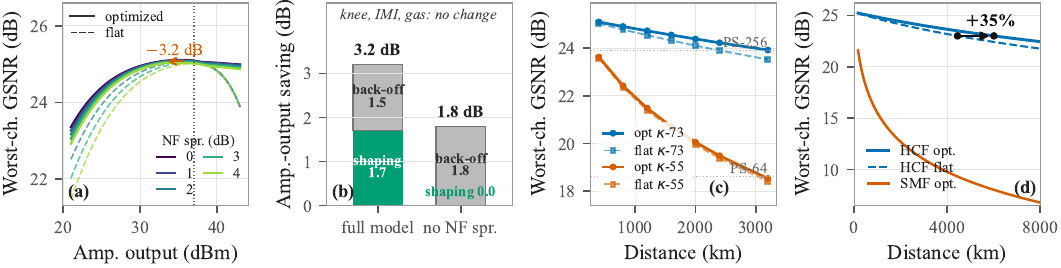}
		\vspace{-.2cm}
		\caption{Amplifier budget and reach: (a)~worst-channel GSNR vs total
			amplifier output, 0--4 dB NF spreads; (b)~the saving split into what shaping buys and
			what backing the budget off a flat-topped curve gives, with and without
			the NF spread; (c)~GSNR vs distance, state-of-the-art and
			legacy IMI; (d)~fixed-30 dBm HCF and SMF reach with the 23 dB HCF
			extension.}
		\label{fig:budgetreach}
		\vspace{-.55cm}
	\end{figure*}

	\vspace{-.05cm}

	\emph{Gain Scaling with Spectral Deficits.}
	Figure~\ref{fig:decoupling}(b) sweeps the NF spread, HCF at a 24-dBm
	(deployed-class) budget, SMF at GN-optimal loading. With a flat $T_i$ the NF
	spread is the whole spectral deficit, and the worst-channel gain grows with it
	from 0.01~dB at zero spread to 0.98~dB at 4~dB, tracking the
	$\rho_{\mathrm{ASE}}$-compression law: shaping repairs only the ASE share. SMF
	reaches about half that (0.52~dB at 4~dB), its capped total power unable to
	chase the deficit.

	\vspace{-.1cm}

	\emph{Gas-Line Channels.}
	Figure~\ref{fig:gas} evaluates an L-band grid overlapping the CO$_2$ bands near
	1572 and 1606~nm. The channel-averaged excess loss reaches 0.054~dB/km
	(4.3~dB/span); the optimizer gives up to 5~dB to the affected channels,
	recovering 0.73~dB on the worst, as in \cite{Hong2026_TPO}. Gas is thus the
	deficit that most rewards shaping in this model.
	Only the ASE part is recoverable; the residual notch ISI survives any launch
	profile, adding 5.5~dB to the SNR the worst on-line channel requires. That
	channel sits 0.02~dB below the PS-64-QAM requirement under flat loading and
	0.7~dB above it once shaped. Gas-line avoidance or core sealing to reduce gas-ingress remains the
	remedy for the ISI itself.
	\vspace{-.1cm}

  \emph{Amplifier Budget and Reach.}
Figure~\ref{fig:budgetreach}(a) sweeps the total amplifier output at NF spreads
of 0--4~dB. Extra power buys back only ASE, so every curve saturates against the
flat TRx ceiling and the IMI floor. The two families then differ: flat loading
must spend whatever budget it is given, so pigtail Kerr NLI turns its curves
over at an interior maximum, whereas the optimized curves flatten out, because
under $\sum_i p_i \le P_{\mathrm{tot}}$ the optimizer stops buying power once
more of it no longer helps. The flat maximum is therefore the benchmark: at the
2-dB baseline it is 25.07~dB at 36.2~dBm, and shaped loading reaches the same
GSNR, within a 0.02-dB match tolerance, at 33.0~dBm --- 3.2~dB less total
output. Both points lie below the 37-dBm NF knee, and the high-power NF penalty
contributes nothing to the comparison: disabling that penalty leaves the flat
curve unchanged out to 50~dBm, so pigtail NLI, not the knee, sets the maximum.

Figure~\ref{fig:budgetreach}(b) splits that 3.2~dB into two parts. Flat loading
already matches its own peak, within the same tolerance, at 34.7~dBm, so 1.5~dB
can be recovered by turning the amplifiers down with the profile left flat; we
call this the budget back-off, and it needs no per-channel optimization.
Per-channel shaping buys the remaining 1.7~dB, moving the match point from
34.7 to 33.0~dBm. Removing the NF spread sends the shaping term to zero and
leaves the back-off unchanged, so the NF spread is the deficit that shaping
repairs, while the back-off reflects the shape of the curve; the NF knee, IMI
and gas change neither part. The shaping term is $1.0$--$1.7$~dB across three
EDFA NF shapes (three-level inversion, a monotone tilt and a GFF-residual
ripple). Both parts are reductions against a spectrally flat benchmark under
this model, not reductions in an amplifier's rated output.

Figure~\ref{fig:budgetreach}(c) sweeps reach at 30~dBm for two distributed-IMI
coefficients $\kappa$. At the state-of-the-art $\kappa=-73$~dB/km
\cite{li_ofc2026} the shaping gain grows from 0.07~dB at 400~km to 0.40~dB at
3200~km as ASE takes a larger share of the noise, and the worst channel stays
${>}4.3$~dB above the PS-64-QAM requirement throughout. At $\kappa=-55$~dB/km
the IMI floor lowers both curves by 1.5--5~dB and masks the ASE term,
compressing the 800-km gain to 0.07~dB: power optimization and IMI reduction
address different limits and are complementary.

\begin{table}[!t]
	\caption{Worst-channel reach at a 23-dB GSNR target.}
	\label{tab:reachext}
	\centering
	\scriptsize
	\setlength{\tabcolsep}{8pt}
	\begin{tabular}{cccc}
		\toprule
		$P_{\mathrm{tot}}$ (dBm) & Flat (km) & Shaped (km) & Extension \\
		\midrule
		24 & 1220 & 1720 & $1.41\times$ \\
		27 & 2360 & 3290 & $1.39\times$ \\
		30 & 4440 & 6000 & $1.35\times$ \\
		33 & 7730 & 9710 & $1.26\times$ \\
		\bottomrule
	\end{tabular}
	\vspace{-.62cm}
\end{table}

 \emph{Reach Extension at a Fixed Amplifier Budget.}
\label{subsec:reachadv}
The dB-domain gains of Fig.~\ref{fig:budgetreach}(c) understate the distance
value of shaping, because the two links convert dB into km at very different
rates. Over 400--3200~km the 30-dBm HCF worst channel sits close to the flat
TRx ceiling and loses only 0.53~dB per doubling of distance, so a small dB gain
buys a large distance gain, whereas the ASE- and NLI-limited SMF link loses
2.8~dB per doubling, approaching the GN-model $10\log_{10}N_s$ slope at longer
reach. At a 23-dB target, between the PS-64-QAM and PS-256-QAM points, shaping
extends the 30-dBm HCF reach from $4440$ to $6000$~km (about $35\%$); the same
optimization on the nonlinearity-capped SMF link gains at most $6\%$ over its
own 6--19-dB target range at the flat-optimal total plotted in
Fig.~\ref{fig:budgetreach}(d), rising to $15\%$ when that total is
re-optimized jointly. Table~\ref{tab:reachext} lists the flat and
shaped reach at this target, together with the extension factor they give, for
total amplifier budgets of 24--33~dBm. The factor falls as the budget rises,
from $1.41\times$ to $1.26\times$, and \eqref{eq:sensitivity} accounts for the
trend: shaping repairs only the ASE share of the noise that accumulates with
distance, and removes about 30\% of that share at every budget, so its leverage
shrinks as extra power pushes ASE below the power-neutral floor. The extension
is thus a fixed-budget effect: the same hardware reaches further, rather than
delivering more GSNR over the same distance.
These projections extrapolate to multi-thousand-km HCF transmission,
consistent with $>$2000-km reduced-IMI NANF loops \cite{Nespola2021_loop}, the
sparsely repeatered 6660-km transoceanic transmission over 266-km HCF spans of
\cite{Boddeda2026_transoceanic}, whose IMI (${\approx}{-}69$~dB/km) lies between
our two $\kappa$ values, and the ${>}30$-dBm launch of
\cite{Hong2024_HCF_terabit,Sohanpal2026_launch}. Perturbing the long-haul parameters one at a
time moves the absolute reach but leaves the 30-dBm factor at
$1.36$--$1.39\times$ for $\alpha=0.09$--$0.13$~dB/km (6240 down to 3040~km flat),
$\mathrm{NF}_{\min}\le5.5$~dB, extra span loss $\le3$~dB and a $\pm1$-dB TRx
shift; the $\alpha=0.09$ case is a lower bound, its shaped reach exceeding the
search range. Only IMI moves it: $1.20\times$ at $\kappa={-}65$~dB/km,
$1.06\times$ at $-60$ and $1.00\times$ at $-55$. IMI, not ASE, binds long
haul \cite{li_ofc2026}.

	\textbf{Discussion and Conclusion.}
	\label{sec:discussion}
	Per-channel power optimization gives modest GSNR gains in HCF, capped by the
	flat TRx ceiling, but its value is operational: the same performance from a
	smaller amplifier budget, more reach at a fixed one, and part of the gas-line
	penalty recovered --- all from a launch profile that departs slightly
	from flat. Unlike SMF, where XPM and ISRS force a centrally planned profile,
	HCF's negligible cross-channel sensitivity lets each channel be set from its
	own GSNR telemetry, turning the WDM launch profile into nearly independent
	controls governed by $S_i=\rho_{\mathrm{ASE},i}-2\rho_{\mathrm{SPM},i}$.

	\begin{backmatter}
		\bmsection{Disclosures}The authors declare no conflicts of interest.
		
		\vspace{-.18cm}
		\bmsection{Data Availability}Data underlying the results are available from the authors upon reasonable request.
		\vspace{-.13cm}
	\end{backmatter}

	\bibliography{refs}

	\bibliographyfullrefs{refs}

\end{document}